\documentclass[aps,twocolumn,groupedaddress,prd,showpacs]{revtex4}

\usepackage{bm}                                                           
\usepackage{amsmath}

\begin{document}

\title{Contribution of $\alpha^2$-terms to the total interaction cross
  sections of relativistic elementary atoms with atoms of matter}

\author{L.Afanasyev} \altaffiliation[Also at: ]{CERN EP Division,
  CH-1211 Geneva 23, Switzerland}
\email{Leonid.Afanasev@cern.ch} \author{A.Tarasov} \thanks{Current
  address: Institute for Theoretical Phisics, University of
  Heidelberg, Philosophenweg 19, D-69120, Heidelberg, Germany}
\email{Alexander.Tarasov@mpi-hd.mpg.de} \author{O.Voskresenskaya}
\email{voskr@cv.jinr.ru}
\affiliation{Joint Institute for Nuclear Research, \\
  141980 Dubna, Moscow Region, Russia}

\date{\today}

\begin{abstract}
  It is shown that the corrections of $\alpha^2$ order to the total
  cross sections for interaction of elementary hydrogen-like atoms
  with target atoms, reported in the previously published paper
  [S.Mr\'owczy\'nski, Phys.Rev. {\bf D36}, 1520 (1987)], do not
  include some terms of the same order of magnitude. That results in a
  significant contribution of these corrections in particular cases.
  The full $\alpha^2$-corrections have been derived and it is shown
  that they are really small and could be omitted for most practical
  applications.
\end{abstract}

\pacs{11.80.Fv, 34.50.-s, 36.10.-k}                                            
\maketitle

\section{Introduction}

The experiment DIRAC \cite{dirac}, now under way at PS CERN's, aims to
measure the lifetime of hydrogen-like elementary atoms (EA) consisting
of $\pi^+$ and $\pi^-$ mesons ($A_{2\pi}$) with an accuracy of 10\%.
The interaction of $\pi^+\pi^-$-atoms with matter is of great
importance for the experiment as $A_{2\pi}$ dissociation (ionization)
in such interactions is exploited to observe $A_{2\pi}$ and to measure
its lifetime. In the experiment the ratio between the number of
$\pi^+\pi^-$-pairs from $A_{2\pi}$ dissociation inside a target and
the number of produced atoms will be measured.  The lifetime
measurement is based on the comparison of this experimental value with
its calculated dependence on the lifetime. The accuracy of the cross
sections for interaction of relativistic EA with ordinary atoms, which
are behind all these calculations \cite{AFAN96}, is essential for the
extraction of the lifetime.

Study of interactions of fast hydrogen-like atoms with atoms has a
long history starting from Bethe. One of the resent calculations for
hydrogen and one-electron ions was published in \cite{gill}.
Interactions of various relativistic EA consisting of e$^\pm$,
$\pi^\pm$, $\mu^\pm$, $K^\pm$ were considered in different approaches
\cite{DYLY83,PAK85,MROW86,MROW87,KUPT89,AFAN91a,AFAN93a,AFAN95,HALA99,%
TARA91,VOSK98,AFAN99,IVAN99A}.  In this paper we reconsider
corrections of $\alpha^2$ order to the EA total interaction cross
sections obtained in \cite{MROW87}. (Through this paper $\alpha$ is
the fine-structure constant.)

\section{General formulas}

As shown in \cite{MROW87} analysis of the relativistic EA interaction
with the Coulomb field of target atoms can be performed conveniently
in the rest frame of the projectile EA (anti-lab frame). As the
characteristic transfer momentum is of the order of the EA Bohr momentum,
in this frame after the interaction EA has a non-relativistic velocity
and thus the initial and final states of EA can be treated in terms of 
non-relativistic quantum mechanics. In this manner the well-known
difficulties of the relativistic treatment of bound states can be get
round.

As in the EA rest frame a target atom moves with the relativistic
velocity, its electromagnetic field is no longer pure Coulomb. It
is described by the 4-vector potential $A_\mu=(A_0,\bm{A})$ with
components related to its rest Coulomb potential $U(r)$:
\begin{equation}
  \label{eq:1}
A_0=\gamma U \qquad \bm{A}=\gamma\bm{\beta}U \;.
\end{equation}
Here $\bm{\beta}=\bm{v}/c$, $\bm{v}$ is the target atom velocity in
the EA rest frame and $\gamma$ is its Lorentz-factor. The time-like
component $A_0$ of the 4-potential interacts with the charges of
the particles forming EA and the space component with their currents.

In this paper we consider only EA consisting of spinless particles
($\pi$, $K$-mesons etc.) which are of interest for the DIRAC
experiment. In the Born approximation the amplitudes of transition
from the initial state $i$ to the final one $f$ due to the interaction
with $A_\mu$ can be written as:
\begin{gather}
  \label{eq:2}
  A_{fi}=U(Q)a_{fi}(\bm{q})\;,\allowdisplaybreaks\\
  \label{eq:3}
  U(Q)=2\int\limits_{0}^{\infty} U(r)\frac{\sin Qr}{Q}rdr\;,
  \allowdisplaybreaks\\
  \label{eq:4}
  a_{fi}(\bm{q})=\rho_{fi}(\bm{q})-\bm{\beta} \bm{j}_{fi}(\bm{q})\;.
\end{gather}

The transition densities $\rho_{fi}(\bm{q})$ and transition currents
$\bm{j}_{fi}(\bm{q})$ are expressed via the the EA wave functions
$\psi_{i}$ and $\psi_{f}$ for the the initial and final states:
\begin{gather}
  \label{eq:5}
  \rho_{fi}(\bm{q})=\int \rho_{fi}(\bm{r})\left(
    e^{i\bm{q}_{1}\bm{r}}-e^{-i\bm{q}_{2}\bm{r}}\right)d^3r\;,
  \allowdisplaybreaks\\
  \label{eq:6}
  \bm{j}_{fi}(\bm{q})=\int \bm{j}_{fi}(\bm{r})\left(
    \frac{\mu}{m_1}e^{i\bm{q}_{1}\bm{r}}+
    \frac{\mu}{m_2}e^{-i\bm{q}_{2}\bm{r}}\right)d^3r\;,
  \allowdisplaybreaks\\
  \label{eq:7}
  \rho_{fi}(\bm{r})=\psi_{f}^{\ast}(\bm{r})\psi_{i}(\bm{r})
  \allowdisplaybreaks\\
  \label{eq:8}
  \bm{j}_{fi}(\bm{r})=\frac{i}{2\mu}\left[
    \psi_{i}(\bm{r})\bm{\nabla}\psi_{f}^{\ast}(\bm{r})-
    \psi_{f}^{\ast}(\bm{r})\bm{\nabla} \psi_{i}(\bm{r})\right] \;.
\end{gather}

The EA wave functions $\psi_{i,f}$ and the binding energies
$\varepsilon_{i,f}$ obey the Schr\"odinger equation
\begin{gather}
  \label{eq:9}
  H \psi_{i,f} =\varepsilon_{i,f} \psi_{i,f}\allowdisplaybreaks\\
  H = -\frac{\Delta}{2\mu} + V(r) \notag
\end{gather}
with the Hamiltonian $H$. It is worth noting that the explicit form of
the potential $V(r)$ of the interaction between the EA components have
no influence on the final result of this paper.

In the above equations $m_{1,2}$ are masses of EA components,
$q=(q_0,\bm{q})$ is the transfer 4-momentum. All other kinematic
variables are related by the following equations:
\begin{gather}
  \bm{q}_1=\frac{\mu}{m_1}\bm{q}\,,\;\;
  \bm{q}_2=\frac{\mu}{m_2}\bm{q}\,,\;\;
  \mu=\frac{m_1m_2}{M}\,,\;\; M=m_1+m_2\,,\notag\\
  q=(q_0,\bm{q})\;, \qquad \bm{q}=(q_L,\bm{q}_T)\;,\notag\\
  q_0=\omega_{fi}+\frac{Q^2}{2M}=\bm{\beta}\bm{q}=\beta q_L \,,\;\;
  \omega_{fi} = \varepsilon_f -\varepsilon_i \;, \label{eq:10}\\
  Q=\sqrt{Q^2}\;, \qquad Q^2=\bm{q}^2 -
  q_0^2=\bm{q}_T^2+q_L^2(1-\beta^2)\;.  \notag
\end{gather}

The differential and integral cross sections of the EA transition
from the initial state $i$ to the final state $f$ due to interaction with the
electromagnetic field of the target atom are related to amplitudes
(\ref{eq:2}):
\begin{align}
  \label{eq:11}
\frac{d\,\sigma_{fi}}{d\,\bm{q}_T} &= \frac{1}{\beta^2}
|A_{fi}(q)|^2 \notag\\
\sigma_{fi} &= \frac{1}{\beta^2}\int|A_{fi}(q)|^2 d^2 \bm{q}_T\;.
\end{align}

Formulae (\ref{eq:2}--\ref{eq:11}) allow to calculate the transition
(partial) cross sections in the Born approximation. But for
applications (for example see \cite{AFAN96}) the total cross sections
of the EA interaction with target atoms are also required. Because the
Born amplitudes of the EA elastic scattering are pure real values, the
optical theorem cannot be used to calculate the total cross sections.
Thus they should be calculated as the sum of all partial cross
sections:
\begin{equation}
  \label{eq:12}
\sigma^{\mathrm{tot}}_i=\sum\limits_{f}\sigma_{fi}\;.
\end{equation}

To get a closed expression for the sum of this infinite series
(the so-called ``sum rule'') the transition amplitudes (\ref{eq:2})
are usually rewriten as:
\begin{equation}
  \label{eq:13}
A_{fi}(q)=\langle f | \widehat{A}(q) | i \rangle \;,
\end{equation}
where the operator $\widehat{A}(q)$ does not contain explicit
dependence on the EA final state variables (for example, its energy
$\varepsilon_{f}$ , see bellow). Then, using the completeness relation
\begin{equation}
  \label{eq:14}
\sum_{f} |f\rangle\langle f| =1 \;,
\end{equation}
we can writte sum (\ref{eq:12}) in the form:
\begin{equation}
  \label{eq:15}
\sigma^{\mathrm{tot}}_i=\frac{1}{\beta^2}\int\langle i
|\widehat A^\ast(q) \widehat{A}(q)|i \rangle d^2  \bm{q}_T \;.
\end{equation}

\section{Simplified approach}

One should take some caution when passing from the exact
expressions (\ref{eq:2}--\ref{eq:10}) for the transition amplitudes,
with explicit dependence on the $\varepsilon_f$ (through the
time-like $q_0$ and longitudinal $q_L$ components of 4-vector $q$), to
the approximate one without such dependence.  Otherwise, it is
possible to obtain a physically improper result as it has happened to
the authors of the paper \cite{MROW87} at deriving of the sum
rules for the total cross section of interaction of ultrarelativistic
EA ($\beta=1$) with target atoms. Below we discuss this problem in detail.

The most essential simplification, that arises in the case of
$\beta=1$ is that $Q^2=\bm{q}_T^2$. Thus $U(Q)=U(\bm{q}_T)$ [see
(\ref{eq:10})] and only $A_{fi}$ in (\ref{eq:2}) depends on
$\varepsilon_f$ through the exponential factors
$\exp{(i\bm{q_1}\bm{r})}$ and $\exp{(-i\bm{q_2}\bm{r})}$ in
(\ref{eq:5}) and (\ref{eq:6})
\begin{equation}
  \label{eq:16}
\bm{q}_{1,2} \bm{r}=\frac{\mu}{m_{1,2}} \bm{q}\bm{r}=
\frac{\mu}{m_{1,2}}(q_L z + \bm{q}_T\bm{r}_T) \;, 
\end{equation}
where $q_L=\omega_{fi}+\bm{q}_T^2/2M$ if $\beta=1$.

Now let us take into account the fact that the typical value of $z$ in these
expressions is of the order of the Bohr radius $r_B=1/\mu\alpha$ and the
typical $q_L\sim\omega_{fi}\sim\mu\alpha^2$, thus the product $q_L z$
is of the order of $\alpha$.  Then it seems natural to neglect the
$q_L$-dependence of $a_{fi}$:
\begin{equation}
  \label{eq:17}
a_{fi}(\bm{q}) \approx a_{fi}(\bm{q}_T) \;.
\end{equation}
and consider this case as the zero order approximation to the problem
\cite{MROW87}.  It corresponds to the choice of the operator
$\widehat{A}$ in the form:
\begin{equation}
  \label{eq:18}
\begin{split}
\widehat A(q) = {} & U(\bm{q}_T)\left[e^{i\bm{q}_{1T}\bm{r}_T} -
  e^{-i\bm{q}_{2T}\bm{r}_T} - \right.\\
& \left.(e^{i\bm{q}_{1T}\bm{r}_T}/m_1 +
 e^{-i\bm{q}_{2T}\bm{r}_T}/m_2) \bm{\beta}\widehat{\bm{p}}\right] \;.
\end{split}
\end{equation}
Here $\widehat{\bm{p}}=-i\bm{\nabla}$ is the momentum operator.

Substituting (\ref{eq:18}) in (\ref{eq:15}) results in the
following sum rules \cite{MROW87}, where the total cross section is
expressed as the sum of the ``electric'' $\sigma^{\mathrm{el}}$ and
``magnetic'' $\sigma^{\mathrm{mag}}$ cross sections:
\begin{gather}
\sigma^{\mathrm{tot}}=\sigma^{\mathrm{el}}+\sigma^{\mathrm{mag}}\;,
  \label{eq:19} \allowdisplaybreaks\\
\sigma^{\mathrm{el}}=\int U^2(\bm{q}_T)M(\bm{q}_T) d^2 \bm{q}_T\;, 
\label{eq:20} \\
M(\bm{q}_T) =2(1-S(\bm{q}_T)) \notag \\
S(\bm{q}_T)=\int |\psi(\bm{r})|^2 e^{i\bm{q}_T\bm{r}}d^3r\;; \notag
  \allowdisplaybreaks\\
\sigma^{\mathrm{mag}}=\int U^2(\bm{q}_T)K(\bm{q}_T)d^2\bm{q}_T\;, 
\label{eq:21}\allowdisplaybreaks\\
K(\bm{q}_T)=\int \left [\frac{1}{\mu^2}+
\frac{2}{m_1 m_2}(e^{i\bm{q}\bm{r}}-1)\right]
\left|\bm{\beta}~\bm{\nabla}~\psi_i(\bm{r})\right|^2 d^3r \;.\notag
\end{gather}

These results reproduce the ones obtained in \cite{MROW87}
and differ from the sum rules used in \cite{AFAN96} by the additional
term $\sigma^{\mathrm{mag}}$. For beginning let us consider its
contribution qualitatively. For this purpose the target atom potential
$U(r)$ can be approximated by the screened Coulomb potential:
\begin{equation}
  \label{eq:22}
U(r)=\frac{Z\alpha}{r}e^{-\lambda r},\quad 
\lambda\sim m_e\alpha Z^{1/3},
\end{equation}
where $m_e$ is the electron mass and $Z$ is the atomic number of the
target. The pure Coulomb wave function can be used for $\psi_i$ (i.e.
the contribution of the strong interaction between the EA components
is neglected, see \cite{amir98}). For the ground state it is written
as:
\begin{equation}
  \label{eq:23}
\psi_i(r)=\frac{\mu\alpha^{3/2}}{\sqrt{\pi}} e^{-\mu\alpha r}\;.
\end{equation}
Under such assumptions for the ground state the following results can
be easily obtained:
\begin{align}
  \label{eq:24}
  \sigma^{\mathrm{el}} &= \frac{8\pi
    Z^2}{\mu^2}\left[\ln\biggl(\frac{2\mu}{Z^{1/3}m_e}\biggl) -
    \frac{3}{4}\right]\; ,\\
  \label{eq:25}
  \sigma^{\mathrm{mag}} &= \frac{4\pi}{3}\left(\frac{Z\alpha
      }{\lambda}\right)^2 +
  O(\alpha^2 \sigma^{\mathrm{el}}) =\notag\\
  &=\frac{4\pi Z^{4/3}\alpha^2}{3 m_e^2} + O(\alpha^2
  \sigma^{\mathrm{el}})\;.
\end{align}

It is seen that in spite of $\alpha^2$ in the numerator of
$\sigma^{\mathrm{mag}}$ the electron mass square in the denominator
makes the contribution of the ``magnetic'' term in (\ref{eq:19}) not
negligible with respect to the ``electric'' one, especially for the
case of EA consisting of heavy hadrons and low $Z$ values.

To obtain exact numerical values we have precisely repeated the
calculations made in \cite{MROW87}.  More accurate presentation of the
target atom potential, namely, the Moli\'ere parametrization of the
Thomas-Fermi potential \cite{mol} was used as in \cite{MROW87}:
\begin{gather} \label{eq:26} 
U(r)=Z\alpha \sum_{i=1}^{3}\frac{c_i e^{-\lambda_i r}}{r}\, ;\\
c_1=0.35,\quad c_2=0.55,\quad c_3=0.1\, ;\notag\\
\lambda_1=0.3\lambda_0,\: \lambda_2=1.2\lambda_0,\: 
\lambda_3=6\lambda_0,\: \lambda_0=m_e \alpha Z^{1/3}/0.885.\notag
\end{gather}
The values of the ``electric'' (el) and ``magnetic'' (mag) total cross
sections (in units of cm$^2$) and their ratio (mag/el) are presented
in Table~\ref{tab:1} for various EA and target materials. The values
published in \cite{MROW87} are given in parentheses. It is seen
that the ``electric'' cross sections coincide within the given
accuracy, but the ``magnetic'' ones are underestimated in
\cite{MROW87}. It is worth noting that the correct values of
$\sigma^{\mathrm{mag}}$ do not depend on EA masses as it follows
from the simplified approximation result (\ref{eq:25}).  The ratio values
confirm the above estimation about the ``magnetic'' term contribution.
Thus, inaccuracy in the calculations did not allow the authors of
\cite{MROW87} to observe such a significant contribution of
$\sigma^{\mathrm{mag}}$ in their results.

\begin{table}[htbp]
    \caption{The ``electric'' (el) and ``magnetic'' (mag) total cross 
      sections in units of cm$^2$ and their ratio (mag/el) in \%
      for EA consisting of $\pi$ and $K$ mesons ($A_{2\pi}$, 
      $A_{\pi K}$, $A_{2K}$) and target materials with the
      atomic number $Z$. The values published in \cite{MROW87} are
      given in parentheses.}
    \label{tab:1}
\begin{ruledtabular}
\begin{tabular*}{\hsize}{r@{\extracolsep{0ptplus1fil}}l%
@{\extracolsep{0ptplus1fil}}l@{\extracolsep{0ptplus1fil}}%
l@{\extracolsep{0ptplus1fil}}l}
Z && \multicolumn{1}{c}{$A_{2\pi}$}& \multicolumn{1}{c}{$A_{\pi K}$} & 
\multicolumn{1}{c}{$A_{2K}$}\\
\colrule\vphantom{$\bigl)$}
 6 & el    &$3.03\cdot10^{-22}$  &$1.37\cdot10^{-22}$  &$3.08\cdot10^{-23}$\\
   &       &$(3.1\cdot10^{-22})$ &$(1.4\cdot10^{-22})$ &$(3.0\cdot10^{-23})$\\
 6 & mag   &$6.73\cdot10^{-24}$  &$6.73\cdot10^{-24}$  &$6.73\cdot10^{-24}$\\
   &       &$(2.5\cdot10^{-24})$ &$(1.3\cdot10^{-24})$ &$(0.3\cdot10^{-24})$\\
 6 & mag/el&2.22\% & 4.90\% & 21.9\%\\                   
\colrule\vphantom{$\bigl)$}                                
13 & el    &$1.33\cdot10^{-21}$  &$6.08\cdot10^{-22}$  &$1.37\cdot10^{-22}$\\
   &       &$(1.3\cdot10^{-21})$ &$(6.2\cdot10^{-22})$ &$(1.4\cdot10^{-22})$\\
13 & mag   &$1.89\cdot10^{-23}$  &$1.89\cdot10^{-23}$  &$1.89\cdot10^{-23}$\\
   &       &$(0.96\cdot10^{-23})$&$(0.55\cdot10^{-23})$&$(0.15\cdot10^{-23})$\\
13 & mag/el&1.41\% & 3.10\% & 13.7\%\\                   
\colrule\vphantom{$\bigl)$}                                
29 & el    &$6.17\cdot10^{-21}$  &$2.84\cdot10^{-21}$  &$6.48\cdot10^{-22}$ \\
   &       &$(6.1\cdot10^{-21})$ &$(2.9\cdot10^{-21})$ &$(6.7\cdot10^{-22})$\\
29 & mag   &$5.50\cdot10^{-23}$  &$5.50\cdot10^{-23}$  &$5.50\cdot10^{-23}$ \\
   &       &$(3.6\cdot10^{-23})$ &$(2.3\cdot10^{-23})$ &$(0.68\cdot10^{-23})$\\
29 & mag/el&0.891\% & 1.94\% & 8.49\%\\                  
\colrule\vphantom{$\bigl)$}                                
47 & el    &$1.55\cdot10^{-20}$  &$7.15\cdot10^{-21}$  &$1.64\cdot10^{-21}$ \\
   &       &$(1.5\cdot10^{-20})$ &$(7.3\cdot10^{-21})$ &$(1.7\cdot10^{-21})$\\
47 & mag   &$1.05\cdot10^{-22}$  &$1.05\cdot10^{-22}$  &$1.05\cdot10^{-22}$ \\
   &       &$(0.79\cdot10^{-22})$&$(0.52\cdot10^{-22})$&$(0.17\cdot10^{-22})$\\
47 & mag/el&0.676\% & 1.46\% & 6.37\%\\                  
\colrule\vphantom{$\bigl)$}                                
82 & el    &$4.46\cdot10^{-20}$  &$2.07\cdot10^{-20}$  &$4.81\cdot10^{-21}$ \\
   &       &$(4.4\cdot10^{-20})$ &$(2.1\cdot10^{-20})$ &$(5.1\cdot10^{-21})$\\
82 & mag   &$2.20\cdot10^{-22}$  &$2.20\cdot10^{-22}$  &$2.20\cdot10^{-22}$ \\
   &       &$(1.9\cdot10^{-22})$ &$(1.3\cdot10^{-22})$ &$(0.48\cdot10^{-22})$\\
82 & mag/el&0.493\% & 1.06\% & 4.58\%\\
\end{tabular*}
\end{ruledtabular}
\end{table}

It is clear that such strong enhancement of the magnetic term in
(\ref{eq:19}) is the consequence of its inverse power dependence
(\ref{eq:25}) on the small screening parameter $\lambda$. It is also
easy to see that the origin of such unnatural dependence is in the
behaviour of the factor $K(\bm{q}_T)$ at small $\bm{q}_T$ in
(\ref{eq:21}).  This factor, contrary to $M(\bm{q}_T)$ in
(\ref{eq:20}), does not approach zero at $\bm{q}_T\to0$. But at
$\beta=1$ such behaviour of $K(\bm{q}_T)$ is in conflict with
some general properties of transition amplitudes (\ref{eq:4}), which
follow from the continuity equation:
\begin{equation}
  \label{eq:27}
  \omega_{fi}\rho_{fi}(\bm{q}) - \bm{q}\bm{j}_{fi}(\bm{q})=0 \;.
\end{equation}
(The latter can be derived from the Schr\"odinger equation
(\ref{eq:9})).  Indeed, rewriting the continuity equation in the form:
\begin{equation}
  \label{eq:28}
  \begin{split}
&\omega_{fi}\rho_{fi}(\bm{q}) - q_L \bm{\beta}\bm{j}_{fi}(\bm{q})-
\bm{q}_T\bm{j}_{fi}(\bm{q})=\\
&\omega_{fi} [\rho_{fi}-\bm{\beta}\bm{j}_{fi}(\bm{q})]-
\bm{q}_T^2 \bm{\beta}\bm{j}_{fi}(\bm{q})/2M - 
\bm{q}_T\bm{j}_{fi}(\bm{q})=0\;,
  \end{split}
\end{equation}
it is easy obtain
\begin{equation}
  \label{eq:29}
  \begin{split}
a_{fi}(\bm{q})&=\rho_{fi}(\bm{q})-\bm{\beta} \bm{j}_{fi}(\bm{q})\\
&=\frac{1}{\omega_{fi}} \bigl[\bm{q}_T^2 \bm{\beta}\bm{j}_{fi}(\bm{q})/2M +  
\bm{q}_T\bm{j}_{fi}(\bm{q})\bigr]\;.
  \end{split}
\end{equation}
Thus all transition amplitudes become zero at $q_T=0$.  Therefore,
any transition cross section (\ref{eq:11}) can depend on
the screening parameter $\lambda$ at least only logarithmically, but
never like inverse power of this parameter.  The same is valid for the
sum (\ref{eq:12}) of this quantities, i.e the total cross section.

\section{Accurate formulas}

Since the $\lambda$-dependence of the magnetic term in (\ref{eq:25})
is contradictory to the general result, we must conclude that there is
a fallacy in the deriving of sum rules (\ref{eq:19}) somewhere.  To
understand the origin of the error, made by the authors of \cite{MROW87},
let us go back to quantities (\ref{eq:5}),(\ref{eq:6}) and expand
them in powers of the longitudinal momentum transfer $q_L$:
\begin{align}
\rho_{fi}=\sum_{n=0}^{\infty}\rho_{fi}^{(n)}\;,\quad &
\rho_{fi}^{(n)}=\frac{q_{L}^{n}}{n!}\biggl(\frac{d^n}{dq_{L}^{n}}\rho_{fi}
\biggl)\biggl\vert_{q_L=0}\;,
  \label{eq:30}\allowdisplaybreaks\\
\bm{j}_{fi}=\sum_{n=0}^{\infty}\bm{j}_{fi}^{(n)}\;,\quad&
\bm{j}_{fi}^{(n)}=\frac{q_{L}^{n}}{n!}\biggl(\frac{d^n}{dq_{L}^{n}}
\bm{j}_{fi} \biggl)\biggl\vert_{q_L=0}\,.
  \label{eq:31}
\end{align}

It is easily shown that terms of these expansions obey the following
estimation:
\begin{equation}
  \label{eq:32}
\rho_{fi}^{(n)}\propto\alpha^n\;,\quad
j_{fi}^{(n)}\propto\alpha^{n+1}\;.
\end{equation}
The additional power of $\alpha$ in the current expansion coefficients,
in comparison with the density one, reflects the ordinary relation
between the values of current and density in the hydrogen-like atoms.

Expanding (\ref{eq:4}) and taking into account (\ref{eq:32}) it
seems reasonable to group terms with the same order of $\alpha$ rather
than $q_L$ as was done in \cite{MROW87}. Then the successive terms
of the $a_{fi}$ expansion in powers of $\alpha$ are
\begin{equation}
\begin{split}
  \label{eq:33}
a_{fi} &= \sum_{n} a_{fi}^{(n)} \\
a_{fi}^{(n)} &= \rho_{fi}^{(n)} - \bm{\beta}\bm{j}_{fi}^{(n-1)}\;.
\end{split}
\end{equation}

From above it is clear that in the ``natural'' approximation
(\ref{eq:17}) includes $a_{fi}^{(0)}$ and only one part of the
term $a_{fi}^{(1)}$ of expansion (\ref{eq:33}), namely:
\begin{equation}
  \label{eq:34}
  \bm{\beta}\bm{j}_{fi}^{(0)}=-\frac{i}{\mu}
  \int \psi_f^\ast E(\bm{q}_T,\bm{r}_T)
  \frac{\partial \psi_i}{\partial z} d^3 r \;,
\end{equation}
while the second one
\begin{equation}
  \label{eq:35}
\rho_{fi}^{(1)}=i q_L \int \psi_f E(\bm{q}_T,\bm{r}_T) z \psi_i d^3 r
\end{equation}
was omitted according to the reasoning of approximation
(\ref{eq:17}).  In equations (\ref{eq:34}), (\ref{eq:35})
$E(\bm{q}_T,\bm{r}_T)$ denotes:
\begin{equation}
  \label{eq:36}
E(\bm{q}_T,\bm{r}_T) = \frac{\mu}{m_1}e^{i\bm{q}_{1T}\bm{r}_T} +
\frac{\mu}{m_2}e^{-i\bm{q}_{2T}\bm{r}_T} \;.
\end{equation}

Let us consider this neglected part in detail. As it is proportional
to $q_L=\omega_{fi}+\bm{q}_T^2/2M$ and therefore explicitly depends on
$\varepsilon_f$, one cannot use completeness relation (\ref{eq:14}) to
calculate its contribution to the total cross section directly. First
we need to transform it to the form free of such dependence.  It can
be done with the help of Schr\"odinger equation (\ref{eq:9}).
\begin{multline}
  \label{eq:37}
\varepsilon_{fi} \int \psi_f^\ast(\bm{r}) E(\bm{q}_T,\bm{r}_T) z
\psi_i(\bm{r}) d^3 r = \allowdisplaybreaks\\
\int \psi_f^\ast(\bm{r}) \left\{\varepsilon_f E(\bm{q}_T,\bm{r}_T) z -
\varepsilon_i E(\bm{q}_T,\bm{r}_T) z\right\} \psi_i(\bm{r}) d^3 r=
\allowdisplaybreaks\\
\int \psi_f^\ast(\bm{r}) [H, E(\bm{q}_T,\bm{r}_T)z]\psi_i(\bm{r}) d^3 r
\end{multline}
The commutator in this relation is easily calculated and after simple
algebra we get the following result:
\begin{gather}
  \label{eq:38}
\rho_{fi}^{(1)}(\bm{q}) =  -\frac{i}{\mu} 
\int \psi_f^\ast(\bm{r}) E(\bm{q}_T,\bm{r}_T)
\frac{\partial\psi_i(\bm{r})}{\partial z} d^3 r
 + \Delta\rho_{fi}^{(1)}(\bm{q})\allowdisplaybreaks\\
\begin{split}
&\Delta\rho_{fi}^{(1)}(\bm{q})= \\
&i \int \psi_f^\ast(r) \left[\frac{\mu}{m_1} e^{i\bm{q}_{1T}\bm{r}_T}
\widehat{O}_1 + \frac{\mu}{m_2} e^{-i\bm{q}_{2T}\bm{r}_T}\widehat{O}_2
\right ] z \psi_i(r) d^3 r \label{eq:39}
\end{split}\allowdisplaybreaks\\
\widehat{O}_{1,2}=\frac{\bm{q}_T^2 \pm 2 \bm{q}_T
  \widehat{\bm{p}}}{2m_{1,2}}\;,\quad \widehat{\bm{p}}=-i\bm{\nabla}\;.
  \label{eq:40}
\end{gather}

It is seen that ``large'' (nonvanishing at $\bm{q}_T=0$) parts of two
terms (\ref{eq:34}) and (\ref{eq:38}), contributing to $a_{fi}^{(1)}$,
are equal and opposite in sign, so that in the resulting expression
they cancel each other, leaving only the term with the ``correct''
behaviour at small $\bm{q}_T$:
\begin{equation}
  \label{eq:41}
a_{fi}^{(1)}= \Delta\rho_{fi}^{(1)}(\bm{q})
\end{equation}

The same is valid for any $a_{fi}^{(n)}$. Applying Schr\"o\-dinger
equation (\ref{eq:9}) to reduce by one the power of $q_L$ in the
expression:
\begin{multline}
  \label{eq:42}
  \rho_{fi}^{(n)}(\bm{q})= \frac{(i q_L)^n}{n!} 
  \int \psi_f^\ast(r) \left[\left(\frac{\mu}{m_1}\right)^n
    e^{i\bm{q}_{1T}\bm{r}_T}\right. + \\ + \left. (-1)^{n+1}
    \left(\frac{\mu}{m_2}\right)^n e^{-i\bm{q}_{2T}\bm{r}_T} \right ]
  z^n \psi_i(r) d^3 r\;,
\end{multline}
one can represent it in the form:
\begin{gather}
  \rho_{fi}^{(n)}(\bm{q}) = \bm{\beta}\bm{j}_{fi}^{(n-1)}(\bm{q}) +
  \Delta\rho_{fi}^{(n)}(\bm{q}) \label{eq:43}\allowdisplaybreaks\\
\begin{split}
  \Delta\rho_{fi}^{(n)}(\bm{q}) = & \frac{i(iq_L)^{n-1}}{n!}  \int
  \psi_f^\ast(r) \left[ \left(\frac{\mu}{m_1}\right)^n
    e^{i\bm{q}_{1T}\bm{r}_T}
    \widehat{O}_1 + \right. \\
  & + \left.\left(\frac{\mu}{m_2}\right)^n
    e^{-i\bm{q}_{2T}\bm{r}_T}\widehat{O}_2 \right] z^n \psi_i(r) d^3 r
  \label{eq:44}\;.
\end{split}
\end{gather}
Finally one has
\begin{equation}
  \label{eq:45}
a_{fi}^{(n)}= \Delta\rho_{fi}^{(n)}(\bm{q})\;.
\end{equation}
That confirms the qualitative result (\ref{eq:29}), derived with
help of continuity equation (\ref{eq:27}).

The remaining $\varepsilon_f$-dependence of right-hand side of
(\ref{eq:44}) can be removed by repeatedly applying the Schr\"o\-dinger
equation (\ref{eq:9}), which allows the transition
amplitudes to be represented in the form (\ref{eq:13}).

From $z$-dependence of the integrand in (\ref{eq:44}) it is easy to
derive that $a_{fi}^{(2k)}=0$ for odd $\Delta (lm)_{fi}$,
and $a_{fi}^{(2k+1)}=0$ for even $\Delta (lm)_{fi}$ , where $\Delta
(lm)_{fi}=(l_f-l_i)-(m_f-m_i)$, and $l_i$, $l_f$, $m_i$, $m_f$ are 
the orbital and magnetic quantum numbers of the initial $i$
and final $f$ states (the quantization axis is supposed to be
$z$-axis).  Thus ``odd'' and ``even'' terms of expansion
(\ref{eq:33}) do not interfere and therefore in the expansion of the
$\sigma^{\mathrm{tot}}$ in the powers of $\alpha$
\begin{equation}
  \label{eq:46}
\sigma^{\mathrm{tot}}=\sum_{n=0}^\infty \sigma^{(n)}\;, \quad
\sigma^{(n)}\propto\alpha^n
\end{equation}
only even powers are present.

The structure of the zero order term of this expansion is well
established [see (\ref{eq:20})]. In view of the above discussion one
can be assured that the higher order terms are numerically negligible and
should not be discussed in detail. Nevertheless, for completeness of the
consideration we present the expression for the contribution of the 
$\alpha^2$-term to the total cross section which includes the term 
$\left|a_{fi}^{(1)}\right|^2$ and the interference term
$a_{fi}^{(0)}\,a_{fi}^{(2)}$.

\begin{gather}
  \sigma^{(2)} = - \int U^2(\bm{q}_T) W(\bm{q}_T) d^2\bm{q}_T
  +O(\alpha^4)\;,
\label{eq:47}
\allowdisplaybreaks\\
\begin{split}
  W(\bm{q}_T)=\frac{1}{4 m_1 m_2} \int z^2 &
  \left[ q^4|\psi_i(\bm{r})|^2 -\right.\\
  &\quad-\left.\left|2\bm{q}_T\widehat{\bm{p}}\psi_i(\bm{r})\right|^2
  \right] e^{i\bm{q}\bm{r}}d^3r\;.
\end{split}\notag
\end{gather}

The ``correct'' $q_T$-dependence of the last integrand excludes a
possibility of some extra $\lambda$-dependence arising, which could
dramatically enhance the contribution of this term (as happened to the
$\sigma^{\mathrm{mag}}$ term in \cite{MROW87}). This can be illustrated
by the explicit expression for the case of the screened Coulomb
potential (\ref{eq:22}) and the EA ground state (\ref{eq:23}):
\begin{equation}
\sigma^{(2)}=  - \frac{8\pi(Z\alpha)^2}{5M\mu}
\left[\ln\left(\frac{2\mu}{Z^{1/3}m_e}\right) -\frac{4}{5} \right] \;.
\label{eq:48}
\end{equation}
Because of numerical smallness of $\alpha^2$ this term can be
successfully neglected compared to (\ref{eq:24}).

Thus in most practical applications, in which the required relative
accuracy is less than $10^{-4}$, only the zero order term, that
considers the pure Coulomb interaction and only the transverse transfer
momentum, should be taken into account for calculation of the
relativistic atom-atom cross sections.  This result warrants the usage
of the simple expression:
\begin{equation}
  \label{eq:49}
\sigma^{\mathrm{tot}}=2\int U^2(q_{T})\left[1-S(q_{T})\right] d^2q_{T}
\end{equation}
for the total cross section calculation for the Born approximation in
\cite{AFAN96} and for the Glauber extensions in \cite{AFAN99}.

The authors express their gratitude to professors S.Mr\'ow\-czy\'nski,
L.Nemenov and D.Trautmann for helpful discussions. This work is
partially supported by RFBR grant 01--02--17756.


\begin{thebibliography}{99}
\bibitem{dirac} B. Adeva et al., {\it Lifetime measurement of
    $\pi^+\pi^-$ atoms to test low energy QCD predictions}, Proposal
  to the SPSLC, CERN/SPSLC 95--1, SPSLC/P 284, Geneva, 1995.

\bibitem{AFAN96} L.G.Afanasyev and A.V.Tarasov: Yad.Fiz. {\bf 59},
  2212 (1996); Phys.Atom.Nucl. {\bf 59}, 2130 (1996).

\bibitem{gill}  G.H.Gillespie, Phys.Rev. {\bf A18}, 1967 (1978);\\
G.H.Gillespie, M.Inokuti, Phys.Rev. {\bf 22A}, 2430 (1980).

\bibitem{DYLY83} L.S.Dul'yan and A.M.Kotsinyan: Yad.Fiz. {\bf 37}, 137\\
  (1983); Sov.J.Nucl.Phys. {\bf 37}, 78 (1983).

\bibitem{PAK85} A.S. Pak, A.V. Tarasov: JINR-P2-85-903, Dubna 1985.

\bibitem{MROW86} S.Mr\'owczy\'nski: Phys.Rev. {\bf A33}, 1549 (1986).
  
\bibitem{MROW87} S.Mr\'owczy\'nski: Phys.Rev. {\bf D36}, 1520
  (1987);\\K.G.Denisenko and S.Mr\'owczy\'nski: Phys.Rev. {\bf D36},
  1529 (1987).

\bibitem{KUPT89} A.V.Kuptsov, A.S.Pak and S.B.Saakian: Yad.Fiz. {\bf
    50}, 936 (1989); Sov.J.Nucl.Phys. {\bf 50}, 583 (1989).
  
\bibitem{AFAN91a} L.G.Afanasyev: JINR-E2-91-578, Dubna 1991.
  
\bibitem{AFAN93a} L.G.Afanasyev and A.V.Tarasov: JINR E4-93-293,\\
  Dubna, 1993.
  
\bibitem{AFAN95} L.G.Afanasyev: Atomic Data and Nuclear Data Tables,
  {\bf 61}, 31 (1995).

\bibitem{HALA99} Z.Halabuka  et al.: Nucl.Phys. {\bf 554}, 86 (1999).
  
\bibitem{TARA91} A.V.Tarasov and I.U.Christova, JINR P2-91-10, Dubna,
  1991.
  
\bibitem{VOSK98} O.O.Voskresenskaya, S.R.Gevorkyan and A.V.Tarasov:
  Yad.Fiz. {\bf 61}, 1628 (1998); Phys.Atom.Nucl. {\bf 61}, 1517
  (1998).
  
\bibitem{AFAN99} L.Afanasyev A.Tarasov and O.Voskresenskaya: J.Phys.
  {\bf G 25}, B7 (1999).
  
\bibitem{IVAN99A} D.Yu.Ivanov and L.Szymanowski: Eur.Phys. J.{\bf A5},
  117 (1999).
  
\bibitem{amir98} Amirkhanov I. et al., Phys.\ Lett. {\bf B 452}, 155
  (1999).

\bibitem{mol} Moli\`ere~G., {\it Z.~Naturforsch.} {\bf 2A}, 3 (1947).

\end{thebibliography}
\end{document}